\documentclass[preprint,amsmath,amssymb,aps,showkeys,showpacs]{revtex4}
\usepackage[english]{babel}
\usepackage{graphicx}
\usepackage{graphics}
\usepackage{amsmath}
\usepackage{dcolumn}
\usepackage{amssymb}
\usepackage{bm}

\begin{document}
\title{Harmonic oscillator in an elastic medium with a spiral dislocation}
\author{A. V. D. M. Maia}
\affiliation{Departamento de F\'isica, Universidade Federal da Para\'iba, Caixa Postal 5008, 58051-900, Jo\~ao Pessoa, PB, Brazil.}

\author{K. Bakke}
\email{kbakke@fisica.ufpb.br}
\affiliation{Departamento de F\'isica, Universidade Federal da Para\'iba, Caixa Postal 5008, 58051-900, Jo\~ao Pessoa, PB, Brazil.}

\begin{abstract}

We investigate the behaviour of a two-dimensional harmonic oscillator in an elastic medium that possesses a spiral dislocation (an edge dislocation). We show that the Schr\"odinger equation for harmonic oscillator in the presence of a spiral dislocation can be solved analytically. Further, we discuss the effects of this topological defect on the confinement to a hard-wall confining potential. In both cases, we analyse if the effects of the topology of the spiral dislocation gives rise to an Aharonov-Bohm-type effect for bound states.

\end{abstract}


\maketitle

\section{Introduction}

At present days, it is well-known that linear topological defects in solids can be described by differential geometry \cite{kleinert,kat,val}. In a crystal, these kinds of defects are related to the presence of curvature and torsion. Another line of research goes to the context of general relativity, where it deals with linear topological defects in the spacetime \cite{put}. The common point between these areas of physics is the description of these defects made by the Volterra process \cite{kleinert,put,kat,valdir2}. Up to now, a great deal of study has explored quantum system in the presence of disclinations \cite{fur,fur2,fur4,fur5,term} and screw dislocations \cite{valdir2,fur3,fil,hall,spin,shell,bfb} in the context of condensed matter physics. In particular, torsion effects related to the presence of topological defects are in the interests of studies of semiconductors \cite{semi,semi2,semi3,semi4}. Moreover, spacetimes with space-like disclinations and with space-like dislocations \cite{put} have also been investigated as backgrounds of relativistic quantum systems \cite{valdir,bf2,vb,eug2,rel2,vb2,fur6}. On the hand, studies of an edge dislocation through differential geometry and its influence on quantum systems have not been widely explored. In recent years, geometric quantum phases have been discussed in the presence of a spiral dislocation \cite{bf}. In this work, we shall discuss the influence of a spiral dislocation (an edge dislocation) on the harmonic oscillator. We shall show that the Schr\"odinger equation for a spinless particle confined to a harmonic oscillator in the presence of a spiral dislocation can be solved analytically. Further, we extend our discussion to the confinement to a hard-wall confining potential.

The structure of this paper is: in section II, we write the Schr\"odinger equation in the background established by the spiral dislocation, and thus we obtain its exact solutions. Then, we analyse the influence of the spiral dislocation on the energy levels of the harmonic oscillator; in section III, we confine the system to a hard-wall confining potential, and thus, obtain the energy levels of the system; in section IV, we present our conclusions.

\section{Harmonic oscillator in the presence of a spiral dislocation}

Let us consider an elastic medium that contains an edge dislocation, which corresponds to the distortion of a circle into a spiral. This defect is called as a spiral dislocation \cite{val}. This topological defects is described by the line element:
\begin{eqnarray}
ds^{2}=dr^{2}+2\beta\,dr\,d\varphi+\left(\beta^{2}+r^{2}\right)d\varphi^{2}+dz^{2},
\label{1.1}
\end{eqnarray}
where the constant $\beta$ is the parameter related to the dislocation. Next, let us confine a spinless particle to the harmonic oscillator potential ($V\left(r\right)=\frac{1}{2}\,m\,\omega^{2}r^{2}$) in the background given by the spiral dislocation (\ref{1.1}). Therefore, the time-independent Schr\"odinger equation for the harmonic oscillator in the presence of the spiral dislocation is (we shall use the units $\hbar=c=1$)
\begin{eqnarray}
\mathcal{E}\psi&=&-\frac{1}{2m}\left(1+\frac{\beta^{2}}{r^{2}}\right)\frac{\partial^{2}\psi}{\partial r^{2}}-\frac{1}{2m}\left(\frac{1}{r}-\frac{\beta^{2}}{r^{3}}\right)\frac{\partial\psi}{\partial r}+\frac{\beta}{m\,r^{2}}\frac{\partial^{2}\psi}{\partial r\,\partial\varphi}-\frac{1}{2m\,r^{2}}\frac{\partial^{2}\psi}{\partial\varphi^{2}}\nonumber\\
[-2mm]\label{1.2}\\[-2mm]
&-&\frac{1}{2m}\frac{\beta}{r^{3}}\frac{\partial\psi}{\partial\varphi}-\frac{1}{2m}\frac{\partial^{2}\psi}{\partial z^{2}}+\frac{1}{2}\,m\,\omega^{2}r^{2}\psi.\nonumber
\end{eqnarray}
Note that we have written the operator $\hat{p}^{2}=-\nabla^{2}$ in Eq. (\ref{1.2}) in terms of the the Laplace-Beltrami operator  \cite{valdir2,fur,fur2}: $\nabla^{2}=\frac{1}{\sqrt{g}}\,\partial_{i}\left(g^{ij}\,\sqrt{g}\,\partial_{j}\right)$, where $g_{ij}$ is the metric tensor, $g^{ij}$ is the inverse of $g_{ij}$ and $g=\mathrm{det}\left|g_{ij}\right|$. Note that the indices $\left\{i,\,j\right\}$ run over the space coordinates.

As we can see in Eqs. (\ref{1.1}) and (\ref{1.2}), this system has the cylindrical symmetry, thus, a simple way of writing the solution to Eq. (\ref{1.2}) is in terms of the eigenvalues of the $z$-components of the angular momentum and the linear momentum operators as: $\psi\left(r,\,\varphi,\,z\right)=e^{il\varphi+ikz}\,R\left(r\right)$, where $k$ is a constant and $l=0,\pm1,\pm2,\pm3,\pm4\ldots$. In this way, we obtain the radial equation: 
\begin{eqnarray}
\left(1+\frac{\beta^{2}}{r^{2}}\right)R''+\left(\frac{1}{r}-\frac{\beta^{2}}{r^{3}}-i\frac{2\beta\,l}{r^{2}}\right)R'-\frac{l^{2}}{r^{2}}\,R+i\frac{\beta\,l}{r^{3}}R-m^{2}\omega^{2}r^{2}R+\left(2m\mathcal{E}-k^{2}\right)R=0.
\label{1.3}
\end{eqnarray}

A possible solution to Eq. (\ref{1.3}) can be given in the form:
\begin{eqnarray}
R\left(r\right)=\exp\left(i\,l\,\tan^{-1}\left(\frac{r}{\beta}\right)\right)\times f\left(r\right),
\label{1.4}
\end{eqnarray}
where $f\left(r\right)$ is an unknown function. After substituting the radial wave function (\ref{1.4}) into Eq. (\ref{1.3}), we obtain the following equation for the function $f\left(r\right)$:
\begin{eqnarray}
\left(1+\frac{\beta^{2}}{r^{2}}\right)f''+\left(\frac{1}{r}-\frac{\beta^{2}}{r^{3}}\right)f'-\frac{l^{2}}{\left(r^{2}+\beta^{2}\right)}\,f-m^{2}\omega^{2}r^{2}\,f+\left(2m\mathcal{E}-k^{2}\right)f=0.
\label{1.5}
\end{eqnarray}

Next, let us define $x=m\omega\left(r^{2}+\beta^{2}\right)$. Hence, by analysing the behaviour of Eq. (\ref{1.5}) at $x\rightarrow\infty$ and $x\rightarrow0$ \cite{griff,landau}, we can write its solution as 
\begin{eqnarray}
f\left(x\right)=e^{-\frac{x}{2}}\,x^{\frac{\left|l\right|}{2}}\,_{1}F_{1}\left(\frac{\left|l\right|}{2}+\frac{1}{2}-\lambda,\,\left|l\right|+1;\,x\right),
\label{1.9}
\end{eqnarray}
where $\,_{1}F_{1}\left(\frac{\left|l\right|}{2}+\frac{1}{2}-\lambda,\,\left|l\right|+1;\,x\right)$ is the confluent hypergeometric function \cite{arf,abra} and $\lambda=\frac{1}{4m\omega}\left(2m\mathcal{E}-k^{2}+m^{2}\omega^{2}\beta^{2}\right)$. Observe that the asymptotic behaviour of a confluent hypergeometric function for large values of its argument is given by \cite{abra}
\begin{eqnarray}
\,_{1}F_{1}\left(a,\,b\,;x\right)\approx\frac{\Gamma\left(b\right)}{\Gamma\left(a\right)}\,e^{x}\,x^{a-b}\left[1+\mathcal{O}\left(\left|x\right|^{-1}\right)\right],
\label{1.11a}
\end{eqnarray}
therefore, it diverges when $x\rightarrow\infty$. With the purpose of obtaining bound states solutions to the Schr\"odinger equation, we need to impose that $a=-n$ ($n=0,1,2,3,\ldots$), i.e., we need that $\frac{\left|l\right|}{2}+\frac{1}{2}-\lambda=-n$. With this condition, the confluent hypergeometric function becomes well-behaved when $x\rightarrow\infty$. Then, from the relation $\frac{\left|l\right|}{2}+\frac{1}{2}-\lambda=-n$, we obtain
\begin{eqnarray}
\mathcal{E}_{n,\,l}=\omega\left(2n+\left|l\right|+1\right)-\frac{1}{2}\,m\,\omega^{2}\beta^{2}+\frac{k^{2}}{2m}.
\label{1.10}
\end{eqnarray}

Hence, Eq. (\ref{1.10}) give us the energy levels of the harmonic oscillator in the presence of a spiral dislocation. We can observe that the topological effects of the spiral dislocation modify the spectrum of energy of the two-dimensional harmonic oscillator, by yielding a new contribution given by the second term of the right-hand side of Eq. (\ref{1.10}). In contrast to the results obtained in Ref. \cite{fur2}, where the presence of the screw dislocation modifies the angular momentum \footnote{It yields an effective angular momentum given by $l_{\mathrm{eff}}=l-\chi\,p_{z}$, where $\chi$ is the parameter that describes the screw dislocation and $p_{z}$ is the eigenvalues of the operator $\hat{p}_{z}=-i\hbar\,\partial_{z}$. Then, the energy levels of the harmonic oscillator are determined by $l_{\mathrm{eff}}$.} and gives rise to an Aharonov-Bohm-type effect for bound states \cite{pesk}, the presence of the spiral dislocation (an edge dislocation) gives no contribution to the angular momentum quantum number in the energy levels. In this sense, there is no Aharonov-Bohm-type effect for bound states \cite{pesk,valdir,fur2}. Note that, by taking $\beta=0$ in Eq. (\ref{1.10}), we recover the spectrum of energy of the two dimensional harmonic oscillator in the absence of defect \cite{fur2}.

\section{hard-wall confining potential}

Let us consider the presence of a hard-wall confining potential in the system discussed in the previous section. Then, for a fixed radius $r_{0}$, due to the presence of the spiral dislocation we have $x_{0}=m\omega\left[r^{2}_{0}+\beta^{2}\right]$. Therefore, the confinement of the quantum particle to a hard-wall potential gives the following boundary condition:
\begin{eqnarray}
f\left(r_{0}\right)\rightarrow f\left(x_{0}\right)=0.
\label{2.1}
\end{eqnarray} 

From Eq. (\ref{1.9}), we have the parameters of the confluent hypergeometric function: $a=\frac{\left|l\right|}{2}+\frac{1}{2}-\lambda$ and $b=\left|l\right|+1$; thus, by considering a fixed value for the parameter $b$ and the parameter $\lambda$ to be large, we can write the confluent hypergeometric function for a fixed $x_{0}=m\omega\left[r^{2}_{0}+\beta^{2}\right]$ as \cite{abra}:
\begin{eqnarray}
_{1}F_{1}\left(a,\,b;\,x_{0}\right)\propto\cos\left(\sqrt{2b\,x_{0}-4a\,x_{0}}-b\frac{\pi}{2}+\frac{\pi}{4}\right).
\label{2.2}
\end{eqnarray}

Hence, from Eq. (\ref{2.1}), we obtain the energy levels that stem from the interaction of quantum particle with the harmonic oscillator subject to a hard-wall confining potential:
\begin{eqnarray}
\mathcal{E}_{n,\,l}=\frac{1}{2m\left(r^{2}_{0}+\beta^{2}\right)}\left(n\pi+\frac{\left|l\right|}{2}\,\pi+\frac{3\pi}{4}\right)^{2}-\frac{1}{2}\,m\,\omega^{2}\beta^{2}+\frac{k^{2}}{2m}.
\label{2.3}
\end{eqnarray}

In Eq. (\ref{2.3}), we have two contributions to the energy levels that arises from the topological effects of the spiral dislocation. The first one corresponds to the presence of an effective radius given by $\rho_{0}=\sqrt{r^{2}_{0}+\beta^{2}}$. The second contribution is given by the second term of the right-hand side of Eq. (\ref{2.3}). By comparing with the Aharonov-Bohm-type effect for bound states discussed in Refs. \cite{vb,vb2,bfb} (where the topology of the screw dislocation also modifies the angular momentum), we have that the effects of the topology of the spiral dislocation on the energy levels (\ref{2.3}) yield no change in the angular momentum quantum number. In this sense, there is no Aharonov-Bohm-type effect for bound states \cite{pesk,valdir,fur2}.

\section{Conclusions}

We have analysed the interaction of a quantum particle with the two-dimensional harmonic oscillator in an elastic medium that possesses a spiral dislocation. We have seen that the effects of the topology of the spiral dislocation yields a new contribution to the spectrum of energy of the harmonic oscillator. By comparing the effects that stems from the presence of the spiral dislocation with the screw dislocation discussed in Ref. \cite{fur2}, we have seen that no change in the angular momentum is given by the topological effects of the spiral dislocation, therefore, no Aharonov-Bohm-type effect for bound states \cite{pesk} can be observed.

We also have analysed the interaction of quantum particle with the harmonic oscillator subject to a hard-wall confining potential in the presence of the spiral dislocation. Two contributions to the energy levels arises from the topological effects of the spiral dislocation, where one of them is given by the presence of a fixed effective radius, $\rho_{0}=\sqrt{r^{2}_{0}+\beta^{2}}$, in the energy levels. Besides, no contribution to the angular momentum is given by the topology of the spiral dislocation, hence, there is no Aharonov-Bohm-type effect for bound states \cite{pesk}.

\acknowledgments{We would like to thank the Brazilian agencies CNPq and CAPES for financial support.}

\end{document}